\documentstyle[aps,manuscript]{revtex}

\begin{document}
\draft
\preprint{}
\title{GLASSY ROUGHNESS OF A CRYSTALLINE SURFACE UPON A DISORDERED  SUBSTRATE}
\author{D. Cule and Y. Shapir }
\address{Department of Physics and Astronomy\\University of Rochester\\
Rochester, NY 14627}

\date{\today}
\maketitle
\begin{abstract}
 The discrete Gaussian model for the surface of a crystal deposited on a 
disordered substrate is studied by Monte Carlo simulations.  A continuous
transition is found from
a phase with a thermally-induced roughness to a glassy one in which the 
roughness is 
driven by the disorder. 
The behavior of the height-height correlations  
is consistent with the one-step replica symmetry broken solution of
the variational approximation. The results differ from the renormalization
group predictions and from  recent simulations of a 2D vortex-glass model which 
belongs to
the same universality class.  
\end{abstract}
\pacs{05.70.Jk, 64.60.Fr, 64.70.Pf, 74.60.Ge}
\narrowtext

 The surface of a crystal formed by the deposition of particles on a flat 
substrate    is undergoing a roughening transition: At temperatures below 
$T_{r}$, the roughening
transition temperature, the surface is flat. For $T>T_{r}$ the periodic 
potential due to the discrete and uniform size of the
particles is irrelevant and the thermal fluctuations cause the surface to 
roughen \cite {chui,noz}.
The intriguing question of what will be the crystalline surface properties with 
a 
disordered substrate has been addressed recently \cite {tsai}. The 
renormalization group (RG)
analysis
predicts a low-temperature superrough phase with surprisingly uncommon 
equilibrium 
and dynamic \cite {tsai,toner} characteristics. Near the superroughening 
transition 
the 
Hamiltonian 
of the surface is that of the 2D sine-Gordon model with random phases 
\cite{cardy,gold,vil}. 
Another system
which belongs to the same universality class 
(and therefore should follow a similar
behavior near criticality) is the two dimensional array of flux lines with the
magnetic field parallel to the superconducting plane in the presence of random 
pinning \cite {fisher}.  In that 
context
the random sine-Gordon model (RSGM) has been studied by a non-perturbative 
variational (harmonic)
approximation \cite {sha,mez,bou} which allows replica-symmetry breaking with a 
low-temperature phase 
drastically 
different
than that found by the RG approach \cite {kor,giam}. Recent numerical 
simulations in 
the weak 
coupling 
limit of the RSGM \cite {bar} have not observed any signature of a transition in 
the 
equilibrium
properties.  A transition was found, however, in the dynamic response in 
qualitative 
agreement with the dynamic RG \cite{tsai}.

In view of these results we present here the study by  large scale 
Monte Carlo (MC) simulations
of the disordered-substrate surface model. The main conclusion is the
existence of a low-temperature glassy rough phase separated by a continuous 
transition  (at the critical temperature anticipated by the analytic 
calculations) from the thermally-rough high-temperature phase. Height-height 
correlations in the glassy phase are consistent with the predictions of the 
one-step
replica symmetry broken solution obtained from the variational approach
\cite{kor,giam}. Finite-size scaling for the ratio of moments of a "modified
order-parameter" provides a first  rough estimate for the correlation length 
exponent
at the transition. 

The surface configurations are described in terms of height variables $h_i$, 
defined on every lattice point $i$ of the 2D basal plane. The Hamiltonian we 
choose
is that of the discrete Gaussian model:

\begin{equation}
 H = \frac{\kappa}{2}\sum_{<i,j>}^{} (h_{i} - h_{j})^{2}, 
\label{E1}
\end{equation}
\noindent
where $\kappa $ is the surface tension.
In the case of a
flat surface the $h_i$ takes integer values in the units of the lattice
spacing $a$ in the direction perpendicular to the surface.
To simulate the disordered substrate
we first assign a random quenched height $d_i$, chosen uniformly (and
independently for each site), in the interval $(-a/2,+a/2]$. 
The height $h_i$ is then forced to take the values $h_i=d_i+n_{i}a$ where
$n_i$ is any, positive or negative, integer.
   
 The calculation of the partition function for a given realization of the 
disorder consists in summing 
$\exp{\{-\beta \cal H\}}$
over all integers $n_i$. 
The relation
to the RSGM is obtained by using the Poisson summation formula. Introducing
continuous fields $\phi_i$, the partition function may be expressed as:

\begin{equation}
 Z =  \left( \prod_{i}\int_{-\infty}^{+\infty} d\phi_{i} \right)
      \sum_{n_{i}=-\infty}^{+\infty} 
      \exp\left\{ -\frac{\kappa\beta}{2}\sum_{<i,j>}(\phi_{i}-\phi_{j})^{2}
      + \sum_{i}2\pi\imath\left(\phi_{i}-d_{i}\right)n_{i}/a \right\}.
\label{E2}
\end{equation}
\noindent
In the continuum limit $\phi_i$ are then replaced by $\phi(\vec{x})$ and the 
discrete Laplacian by a
continuous one. In addition, {\it near the critical point} only the
first harmonic in the local periodic ``potential" for $\phi$ is relevant \cite 
{chui}. When 
higher
(irrelevant) harmonics are discarded the Hamiltonian of the RSGM is obtained: 

\begin{equation}
 Z =  \int_{-\infty}^{+\infty} d\phi(\vec{x}) 
      \exp \left\{ -\int d\vec{x} \left[   
                \frac{\kappa\beta}{2}\left[\nabla\phi(\vec{x}) \right]^{2}
 - \lambda\cos\left( 2\pi\left[\phi(\vec{x}) - d(\vec{x})\right]/a\right)
\right] \right\}.
\label{E3}
\end{equation}
\noindent

The equilibrium scaling properties are manifested in the behavior of the 
height-height correlation function: 

\begin{equation}
C({\bf r}) \equiv \left[\left<
          \overline{ \left( h({\bf r + \bf r_{0}}) - h({\bf r_{0}})\right)^{2} }
               \right>_{T} \right]_{av},
\label{E4}
\end{equation}
\noindent
where $\langle \cdots \rangle_{T}$ denotes a thermal (time) average for a
given realization of disorder and $[\cdots ]_{av}$ represents the
configurational average over different realizations of the disorder.
Overbar denotes the averages over all origins of $\bf r_{0}$ and all directions.

We briefly recall the major results previously obtained for 
the $C({\bf r})$:
Analytically the {\em Renormalization group} \cite {cardy} and the {\em 
variational} \cite {kor,giam} 
calculations
both predict a transition  at the same  critical temperature $T_c=\kappa/\pi$.
  Both approaches also agree on the properties of the 
high temperature phase in which the
discrete nature of the particle is irrelevant and the behavior is that of the 
simple
Gaussian model [without the cosine term in the Hamiltonian of Eq. (\ref{E3})].
 The behavior of $C({\bf r})$ is therefore given by:

\begin{equation}
C({\bf r}) =  \frac{T}{\kappa\pi}\;\ln|{\bf r}|.   
\label{E5}
\end{equation}
\noindent 

As mentioned above the two approaches diverge in their predictions regarding the
behavior for $T<T_c$. RG  calculations \cite {tsai,toner,gold,vil} predict 
a new term with $(\ln|{\bf r}|)^{2}$ which 
will dominate 
at large distances. More precisely the first-order calculations yield the 
following
behavior: 

\begin{equation}
C({\bf r}) =  A C_{0}({\bf r}) + B \tau^{2} ( \ln|{\bf r}|)^{2},   
\label{E6}
\end{equation}
\noindent 
where $\tau \equiv 1-T/T_{c}$, $C_{0}({\bf r}) = T/(\pi\kappa) \ln |{\bf r}|$ is
the correlation function (\ref{E5}), $A$ is a nonuniversal constant,
and $B =2/\pi^{2} + O(\tau)$ is a universal constant.

The variational approach \cite {kor,giam}, on the other hand, predicts the same 
scaling 
in $\ln|\bf r|$ as for $T>T_c$.  
The prefactor, however, becomes $T$-independent and
sticks to its
value at $T_c$ for the whole $T<T_c$ phase:

\begin{equation}
C({\bf r}) = \frac{T_{c}}{\pi\kappa} \ln|{\bf r}|. 
\label{E7}
\end{equation} 

The MC calculations were performed using the Metropolis algorithm
for the discrete Gaussian Hamiltonian Eq. (\ref{E1}) on a 
square lattice with periodic
boundary conditions. The simulations were carried on the CM computers of the
Thinking Machines Corporation. At every time step all the variables $h_{i}$ of
one sublattice were simultaneously updated by increasing or decreasing them
(independently) by one unit. The moves are then accepted or rejected according 
to the Metropolis rules.

Following the approach introduced by Young in spin-glass 
MC \cite {young1,bhattyoung}, for every
realization of disorder two replicas of the system were simulated.
They had random initial conditions (realized by adding random integer
heights on top of the random substrate) and each had its own independent
time evolution. In addition to the information extracted from their overlap
(see below) the two-replica approach allows close monitoring of the 
approach to equilibrium.
After equilibration was established, the measurements were taken 
over a time interval which was one or several 
equilibration times long. 
Depending on the lattice size and the length of the MC runs, the disorder
averaging were performed using 100 to several thousands realizations
of the disorder.

The Monte-Carlo simulations of the static height-height correlation function
of the pure discrete Gaussian model were performed by
 W. Shugard et al \cite{shugard}.
They confirmed the existence of the phase transition 
between a high-temperature rough phase 
characterized by a logarithmic behavior of $C({\bf r})$ and a flat 
low-temperature
phase. We reproduced their results as a special case of our model
without disorder. However, introducing the disorder 
drastically changes the character of the low-temperature phase.

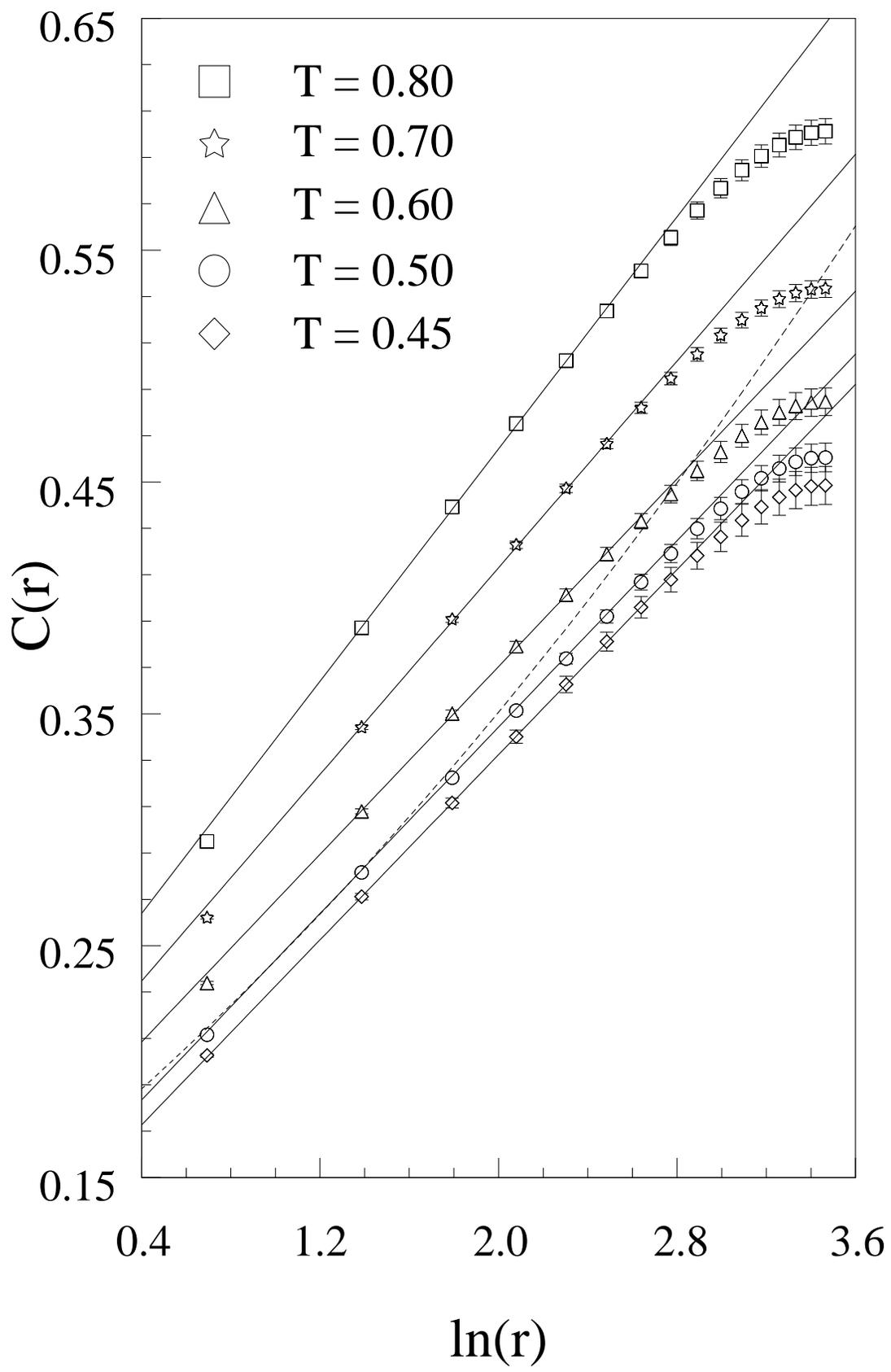
\begin{figure}
\caption{
Semilog plot of $C({\bf r})$ for $L=64$. 
The straight lines are the
best fits to $C({\bf r}) = a(T) + b(T)\:\ln|{\bf r}|$.
The broken line is the RG prediction Eq. (\protect\ref{E6}) at $T=0.5$  with 
$A=1$.
}
\label{F1}
\end{figure}

Fig. \ref{F1} shows on a semilog plot the behavior of $C({\bf r})$ around the 
critical 
temperature $T_{c} = \kappa/\pi = 0.6366$ for  our choice of $\kappa=2$. 
 The simulations were performed with  maximum lattice size of
$L=64$. 
For each temperature, measurements  were repeated, with
a new set of 100 realizations of the disorder, five to ten times.
The average values of these measurements with corresponding
error bars are shown in Fig. \ref{F1}. 

 We compare our results with both the renormalization group predictions 
Eq. (\ref{E6}) and with the results derived by the variational analysis with 
one-step symmetry breaking scheme Eq. (\ref{E7}). In both cases the theoretical 
results
refer to the large $|{\bf r}|$ behavior of $C({\bf r})$ while numerical
results are always limited by the lattice size $L$ and by finite-size effects.

  According to the RG (\ref{E6}), for $T$ near 
$T_{c}$ and for large $|{\bf r}|$ the effect of the second term, 
$B\tau^{2}(\ln|{\bf r}|)^{2}$, should dominate.
To compare with Eq. (\ref{E6}) we need to be close to $T_c$ but not too
close (since the coefficient is proportional to $\tau^{2}$ and the
distance $|{\bf r}|$ at which this term will dominate will be beyond
the size of the system). We therefore choose $T=0.5$ for the comparison.
The broken line shows Eq. (\ref{E6}) for that temperature and $A=1$.
It can be seen clearly that the upper
bending trend of the broken line is inconsistent with the MC data. The
down bending of the MC points for every temperature (which was also observed
in the data taken for the pure system) are due to the finite-size effects.

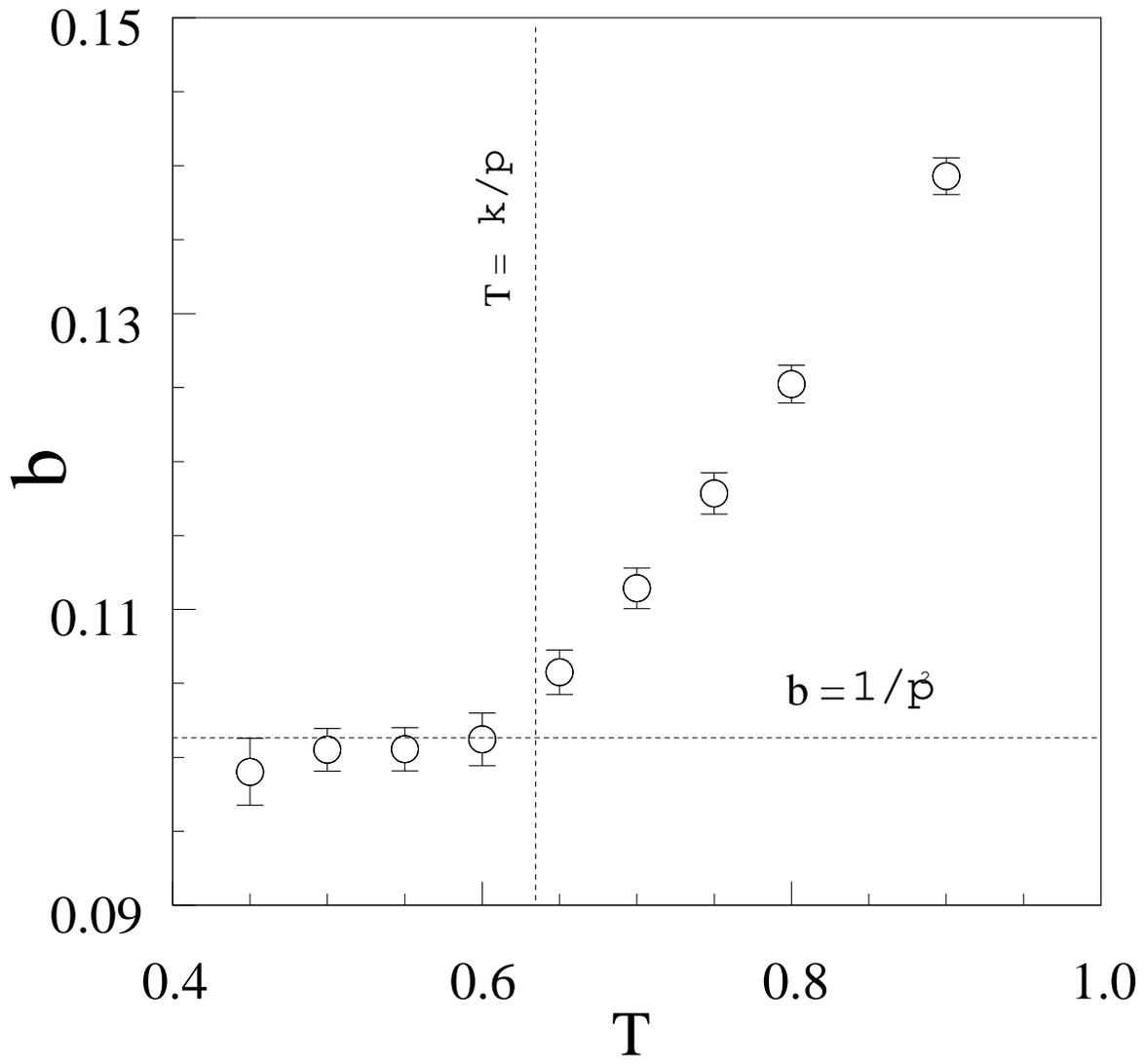
\begin{figure}
\caption{
Plot of the coefficient $b(T)$ from the fitting equation
$C({\bf r}) = a(T) + b(T)\:\ln|{\bf r}|$. 
The vertical dashed line is the analytic $T_{c}$.
The horizontal line is 
the slope predicted by the Eq. (\protect\ref{E7}) for all 
$T\leq T_{c}$. 
}
\label{F2}
\end{figure}

The replica variational approach predicts  (see Eq. (\ref{E7}))
$C({\bf r})$  to remain logarithmic for all $T$ with a $T$-independent 
coefficient for  $T \leq T_{c}$.
The data shown in Fig. \ref{F1} 
is consistent with this behavior. In
our fit we used the values of the $C({\bf r})$ for
$|{\bf r}|$ between 4 and 14 lattice spacings. 
We neglected the higher values of $C({\bf r})$ 
because of the presence of the finite size effects as well as of strong sample
to sample fluctuations. The full straight lines in 
Fig. \ref{F1} are the best fit
curves. In Fig. \ref{F2} we show the temperature dependence of the slopes
for nine values of $T$ between 0.45 and 0.9 including those in Fig. \ref{F1}.
The vertical doted line $T=\kappa/\pi$ is the analytic result for $T_{c}$.
While in high-$T$ phase the slope of $C({\bf r})$
changes linearly with $T$, for the low-$T$ phase it saturates around the value
$1/\pi^{2}$ as is predicted by  Eq. (\ref{E7}).

To gain more insight in the transition and the properties of the 
low-temperature phase, a glassy order-parameter, its correlations, and/or 
probability distribution, should be invoked. 
We tried to look at the local autocorrelation function:

\begin{equation}
q^{\alpha\beta}_{i}(t) = 
\left\{
         \left[ h^{\alpha}_{i}(t_{0}+t)-\overline{h^{\alpha}(t_{0}+t)} \right]
         \left[ h^{\beta}_{i}(t_{0} +\epsilon_{\alpha\beta}t )
               -\overline{h^{\beta}(t_{0} +\epsilon_{\alpha\beta}t )} \right]
\right\},
\label{E8}
\end{equation}
\noindent
where $t_{0}$ is an initial time (larger than the
time required to equilibrate the system) and overbar means average over 
lattice sites. The replica indices $\alpha,\beta =$ take values 1 or 2 
and $\epsilon_{\alpha\beta} = 0$ if $\alpha = \beta$ or
1 if $\alpha \neq \beta$. This local quantity is first averaged over all sites:
 
\begin{equation}
q_{\alpha\beta}(t) = 
\frac{1}{N}\sum_{i=1}^{N}q^{\alpha\beta}_{i}(t),
\label{E9}
\end{equation}
\noindent
$N=L\times L$. This order-parameter definition includes the equal time 
overlap
between different replicas as well as the auto-overlap of the same replica
at different times. In the limit of infinite time separation in the latter the 
probability
distribution for both should coincide. 
It is Gaussian for $T > T_{c}$ but
is expected to deviate from it for $T < T_{c}$. One can then look for the 
phase  transition
by calculating the ratio of the moments

\begin{equation}
g(T,L) = \frac{1}{2}
 \left\{3 - \frac{[\langle q^{4}_{\alpha\beta}\rangle_{T}]_{av}}
          {[\langle q^{2}_{\alpha\beta}\rangle_{T}]_{av}^{2} } \right\}
\label{E10}
\end{equation}
\noindent
after thermal equilibrium has been reached. If the transition is of the second
order the expected finite size scaling of this quantity is
$g(T,L) \sim \tilde{g}(L^{1/\nu}(T-T_{c}))$ where $\nu$ is
the correlation length exponent, $\xi \sim |T-T_{c}|^{-\nu}$.

 Trying to evaluate $q_{\alpha\beta}$ and $g$ we could not reach enough
accuracy to extract reliable results within the computer time available.
A similar problem has been observed in simulations of the $3D$ Ising
spin glass  \cite {bhattyoung} and the $3D$ gauge glass \cite {reger}. One 
possible approach is to
look for a quantity with similar scaling
but for which the statistical errors are smaller. The problem we have to
overcome is that the deviations from the Gaussian distribution occur first
(for $T$ just below $T_{c}$) at very small values of $q$ compared with
its rms. To accentuate the contribution from these small values
we tried to evaluate the ``renormalized" quantity  
$\tilde{q}_{\alpha\beta}(t)$:

\begin{equation}
\tilde{q}_{\alpha\beta} (t)  = 
         \frac{\sum_{i=1}^{N} q_{i}^{\alpha\beta}(t) }
          {\sum_{i=1}^{N}|q_{i}^{\alpha\beta}(t)|}. 
\label{E11}
\end{equation}

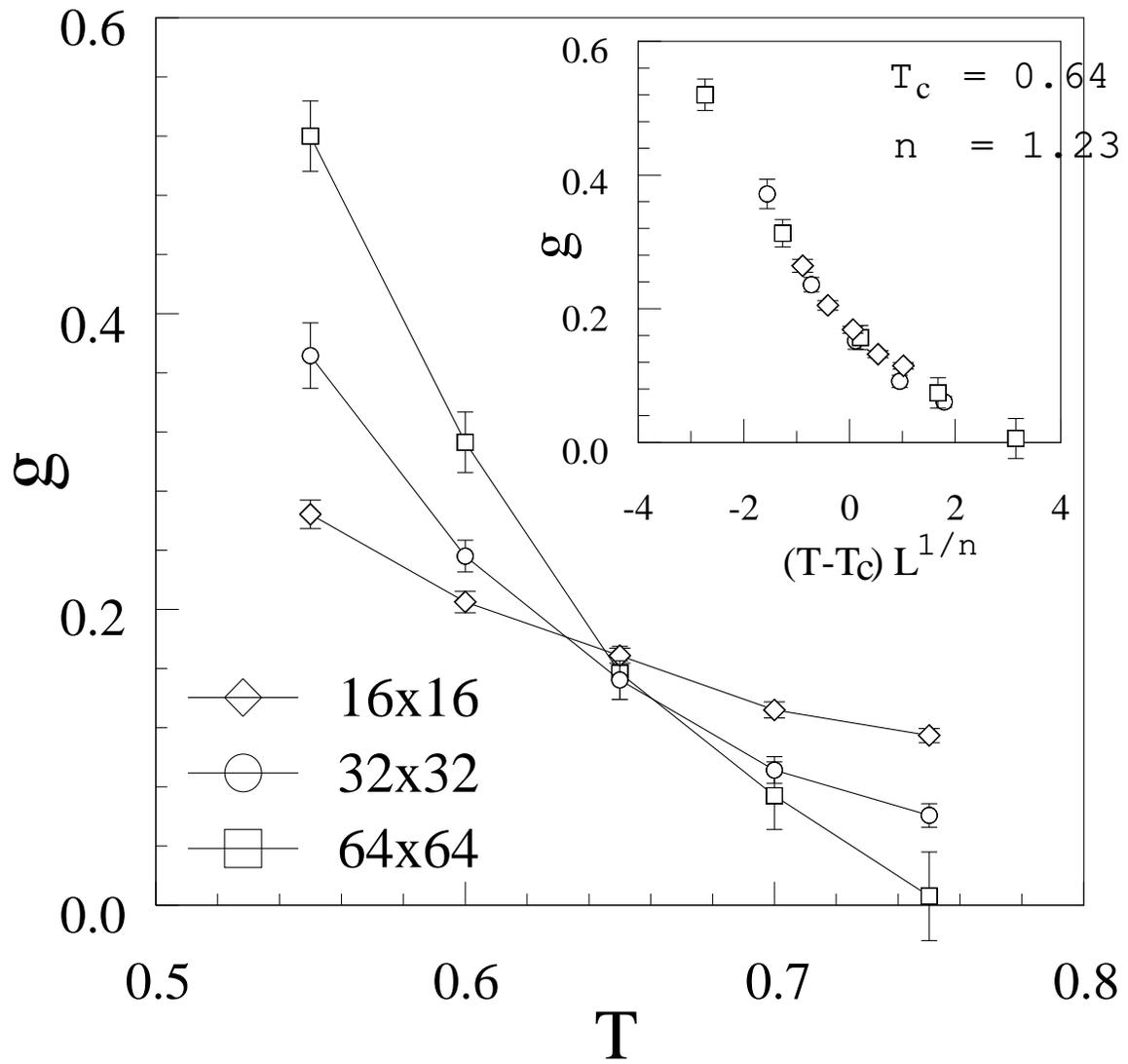
\begin{figure}
\caption{
Plot of $g(T,L)$ $vs$ temperature $T$ for different lattice
sizes. The critical temperature is indicated by the
crossing of the curves. The full line is the guide to the eye.
The inset shows the finite size scaling plot with the indicated values of 
$T_{c}$ and $\nu$ (see text).
}
\label{F3}
\end{figure}

  The distribution $P(\tilde{q})$ (see, e.g., Ref.  \cite {bhattyoung} for its 
definition)  indeed  exhibits a 
transition from a distribution with
one  maximum at $\tilde{q} = 0$  to a distribution with two maxima symmetric 
with
respect to $\tilde{q} = 0$ (which becomes a local minimum). The figures of
$P(\tilde{q})$ will be exhibited in a future publication \cite {culeshapir}. 
Here we only
depict in Fig. \ref{F3} the result of the quantity $g$ defined in Eq. 
(\ref{E10}) but with
$\tilde{q}$ replacing $q$. Equilibration was confirmed by the
convergence of $\tilde{q}_{\alpha\alpha}$ and $\tilde{q}_{\alpha\beta}$.
Depending on the temperature and the size of the simulated system, 
the MC runs for equilibration were between $2^{15}$ to $2^{19}$ MC steps.
For smaller system size the average over disorder is performed using over 1000
samples. However, for large system size, $L=64$, the equilibration time
for temperatures below $T_{c}$ is long, and we worked with a hundred samples.
As in the calculation of $C({\bf r})$, for each temperature the measurements
were repeated several times. Again, the average values of these measurements
with corresponding error bars are shown in Fig. \ref{F3}.
Clearly, Fig. \ref{F3} strongly suggests existence of
the phase transition somewhere in the temperature range 0.63-0.66. 
A finite size scaling plot of $g$ is shown in inset of Fig. \ref{F3}. 
The best fit is
obtained with $T_{c}=0.643 \pm 0.006$ and $ \nu = 1.23 \pm 0.10$. 
The value of $T_{c}$ is in good agreement with the analytic predictions
discussed above.

  We would like to emphasize that the main purpose in presenting the data for
$g$ is to provide an independent confirmation for the existence of the
transition. We believe that replacing $\tilde{q}$ by $q$ will not change the 
critical temperature. In addition, usually the same exponent $\nu$ controls the 
finite-size effects for all thermodynamic quantities, so the 
value found here may be  a first rough estimate for this critical exponent. 
There is no analytic
prediction for $\nu$ within the variational approach
[the RG approach \cite {toner} predicts $\xi \sim\exp (A/\tau^{2})$ and is, 
again, 
in disagreement with our data].
  Our plots of $P(\tilde{q})$ do not exhibit replica symmetry breaking but
they are not accurate enough to rule it out. Simulations of larger systems
should resolve this issue.

  Finally, we comment on the only other numerical work \cite {bar}, in
which no transition was found in the equilibrium $C({\bf r})$: These
simulations were performed in the weak coupling regime $\lambda \ll 1 $
of the DSGM. In contrast  the ones presented here are in the strong coupling
regime with {\it all} harmonics having coefficients O(1). This raises the 
possibility  
that these cases belong to distinct universality classes. Whether such a            
scenario may be reconciled with
the phase transition in the dynamics found in the same weak-coupling simulations 
\cite {bar}, remains to be seen.

 To summarize, numerical evidence  has been presented, from both the
height-height correlations and the overlap distribution, for the
existence of a phase transition in the scaling properties of a
surface upon a disordered substrate. The low-temperature phase has glassy
characteristics. The behavior of $C({\bf r})$ is consistent with that
predicted from the one-step replica symmetry breaking solution although
no direct evidence for (or against) such a breaking was found. Is this symmetry 
broken?
Why there is such a discrepancy between the numerical results and these of the
RG approach which was so successful in explaining the behavior of the pure 
(flat-substrate) surface? These
are two among essential unanswered questions. More studies will be needed to 
reach a fuller understanding of the disordered-substrate surface, the
flux-lines array in $2D$ type-II superconductors, and other physical
systems related to the random-phase sine-Gordon model.

 We are very grateful to T. Giamarchi, T. Hwa, and Y.-C. Tsai for most useful 
discussions.
We also thank Thinking Machines Corporation for use of their
computers.
Acknowledgment is also made to the donors of The Petroleum Research Fund,
administrated by the ACS, for support of this research.

\end{document}